\documentclass[iop,apjl]{emulateapj}
\usepackage{graphicx} % needed for figures
\usepackage{dcolumn} % needed for some tables
\usepackage{bm} % for math
\usepackage{amsfonts,amsmath,amssymb,mathrsfs}
\usepackage{color}
\usepackage{epsfig}
\usepackage{natbib,times}
\usepackage{url}
\usepackage{times}

\shorttitle{HTTP:Unraveling Tarantula's Web}
\shortauthors{Sabbi et al.}

\begin{document}
\title{Hubble Tarantula Treasury Project: Unraveling Tarantula's Web. {\sc i}. Observational overview and first results$^1$}
\author{E. Sabbi\altaffilmark{2}, 
J. Anderson\altaffilmark{2}, 
D. J. Lennon\altaffilmark{3}, 
R. P. van der Marel\altaffilmark{2}, 
A. Aloisi\altaffilmark{2},
M. L. Boyer\altaffilmark{4,5},
M. Cignoni\altaffilmark{6,7},
G. de Marchi\altaffilmark{8},
S. E. de Mink\altaffilmark{2,9,10},
C. J. Evans\altaffilmark{11},
J. S. Gallagher \sc{iii}\altaffilmark{12},
K. Gordon\altaffilmark{2},
D. A. Gouliermis\altaffilmark{13},
E. K. Grebel\altaffilmark{14},
A. M. Koekemoer\altaffilmark{2},
S. S. Larsen\altaffilmark{15},
N. Panagia\altaffilmark{2,16,17},
J. E. Ryon\altaffilmark{12},
L. J. Smith\altaffilmark{18},
M. Tosi\altaffilmark{7},
D. Zaritsky\altaffilmark{19}
}
\email{sabbi@stsci.edu}

\altaffiltext{1}{Based on observations with the NASA/ESA Hubble Space Telescope, obtained at the Space Telescope Science Institute, which is operated by AURA Inc., under NASA contract NAS 5-26555}
\altaffiltext{2}{Space Telescope Science Institute, 3700 San Martin Drive, Baltimore, MD, 21218, USA }
\altaffiltext{3}{ESA - European Space Astronomy Center, Apdo. de Correo 78, 28691Villanueva de la Ca\~{n}ada, Madrid, Spain}
\altaffiltext{4}{Observational Cosmology Lab, Code 665, NASA, Goddard Space Flight Center, Greenbelt, MD, 20771, USA}
\altaffiltext{5}{Oak Ridge Assiciate Universities (ORAU), Oak Ridge, TN 37831, USA}
\altaffiltext{6}{Dipartimento di Astronomia, Universit\`{a} degli Studi di Bologna, via Ranzani, I-40127 Bologna, Italy}
\altaffiltext{7}{Istituto Nazionale di Astrofisica, Osservatorio Astronomico di Bologna, Via Ranzani 1, I-40127 Bologna, Italy}
\altaffiltext{8}{Space Science Department, European Space Agency, Keplerlaan 1, 2200 AG Noordwijk, The Netherlands}
\altaffiltext{9}{Johns Hopkins University, 3400 N. Charles St., Baltimore, MD, 21218, USA}
\altaffiltext{10 }{Hubble Fellow}
\altaffiltext{11}{UK Astronomy Technology Center, Royal Observatory Edinburgh, Blackford Hill, Edinburgh, EH9 3HJ, UK} 
\altaffiltext{12}{Department of Astronomy, University of Wisconsin, 475 North Charter Street, Madison, WI 53706, USA}
\altaffiltext{13}{Universit\"at Heidelberg, Zentrum f\"ur Astronomie, Institut f\"ur Theoretische Astrophysik, Albert-Ueberle-Str.~2, 69120 Heidelberg, Germany}
\altaffiltext{14}{Universit\"at Heidelberg, Zentrum f\"ur Astronomie, Astronomisches Rechen-Institut, M\"onchhofstr. 12-14, 69120 Heidelberg, Germany}
\altaffiltext{15}{Department of Astrophysics / IMAPP, Radboud University Nijmegen, PO Box 9010, 6500 GL Nijmegen, The Netherlands}
\altaffiltext{16}{Istituto Nazionale di Astrofisica, Osservatorio Astrofisico di Catania, Via Santa Sofia 78, 95123 Catania, Italy}
\altaffiltext{17}{Supernova Limited, OYV 131, Northsound Road, Virgin Gorda, British Virgin Islands}
\altaffiltext{18}{ESA/STScI, 3700 San Martin Drive, Baltimore, MD, 21218, USA}
\altaffiltext{19}{Steward Observatory, University of Arizona, 933 North Cherry Avenue, Tucson, AZ 85721, USA}

\begin{abstract}

The Hubble Tarantula Treasury Project (HTTP) is an ongoing panchromatic imaging survey of stellar populations in the Tarantula Nebula in the Large Magellanic Cloud that reaches into the sub-solar mass regime ($< 0.5\, {\rm M_\sun}$).  HTTP utilizes the capability of HST to operate the Advanced Camera for Surveys (ACS) and the Wide Field Camera 3 (WFC3) in parallel to study this remarkable region in the near-ultraviolet, optical, and near-infrared spectral regions, including narrow band H$\alpha$ images. The combination of all these bands provides a unique  multi-band view. The resulting maps of the stellar content of the Tarantula Nebula within its main body provide the basis for investigations of star formation in an environment resembling the extreme conditions found in starburst galaxies and in the early Universe. Access to detailed properties of individual stars allows us to begin to reconstruct the evolution of the stellar skeleton of the Tarantula Nebula over space and time with parcsec-scale resolution. In this first paper we describe the observing strategy, the photometric techniques, and the upcoming data products from this survey and present preliminary results obtained from the analysis of the initial set of near-infrared observations.
\end{abstract}


\section{Introduction}

The Tarantula Nebula  (a.k.a. 30~Doradus, hereafter ``30~Dor'') in the Large Magellanic Cloud (LMC) is one of the most famous objects in astronomy. The first astronomical references to the Tarantula Nebula are more than 150 years old \citep{caille55, herschel47}. For more than a decade the Tarantula's ionizing cluster, R136, was thought to be the most massive known single star  \citep[250-1000 M$_\odot$,][]{feitzinger80}. Apart from a few dissonant voices \citep[e.g.][]{feast53, walborn73, moffat82}, the true nature of R136 continued to elude the majority of the astronomical community until \citet{weigelt85} resolved it into eight components by using holographic speckle interferometry.  

The Tarantula Nebula and R136 are now considered to be the closest (and only known) starburst in the Local Group. The size of the Tarantula Nebula ($\sim 200\, {\rm pc}$ in diameter) and ``local'' density of OB stars parallel those observed in systems characterized by very intense star formation, such as the starburst knots observed in interacting galaxies in the Local Universe and young galaxies at high redshift \citep[$z>5$,][]{meurer97, shapley03, heckman04}. Thus dissecting the stellar populations and inferring an accurate description of the anatomy of the Tarantula Nebula allows us to reconstruct for the first time the temporal and spatial evolution of a prototypical starburst on a sub-cluster scale. 

In this paper we present an introductory overview of the ``Hubble Tarantula Treasury Project'' (HTTP, PI: Sabbi, GO-12939), an ongoing treasury program designed to observe the entire region with the {\it Hubble} Space Telescope (HST) in the near-UV (F275W and F336W), optical (F555W, F658N) and near-IR (F110W and F160W). The program is built on an existing HST monochromatic survey in the F775W filter (GO-12499, PI: Lennon), designed to measure proper motions of runaway candidates. 

\begin{figure*}
\epsscale{1.0}
\plotone{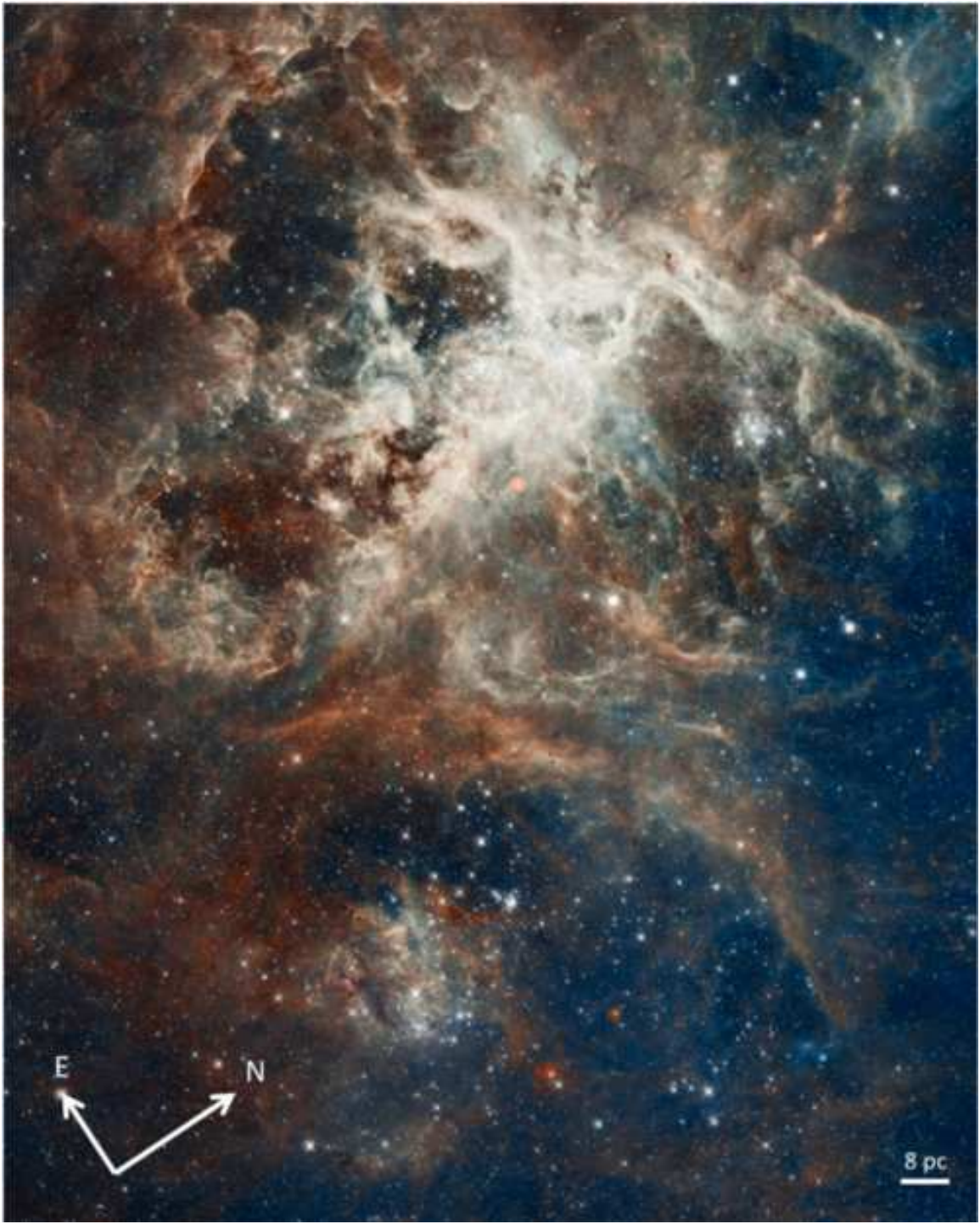}
\caption{\label{f:tarantula} The Tarantula Nebula as seen with HST (Credit NASA, ESA, Lennon; GO-12499). The image is the result of 15 HST pointings with the F775W filter, using WFC3 and ACS in parallel. The projected area in the sky is $\sim 14\arcmin \times12\arcmin$. The dynamic range of the image has been compressed to show  both bright stars and faint filaments of gas. Colors are from ESO WFI exposures in the H$\alpha$ (in red) and S{\sc ii} (in blue) filters. The ionizing cluster R136 is located in the middle of the upper half of the image, in the brightest (white) part of the nebulosity. The small bright cluster to the right of R136 is Hodge 301, and the loose stellar association in the middle of the lower half of the image is NGC~2060. Once observations are completed, HTTP will cover approximately the same region in six additional filters, namely F275W, F336W, F555W, F658N, F110W and F160W.}
\end{figure*}

Bias-subtracted and flat-fielded images, as well as all the raw data, are immediately available to the public and can be downloaded from the Mikulski Archive for Space Telescopes (MAST). Photometric catalogs and mosaicked images for all the filters will be distributed by our team to the community in a timely fashion. The survey progress can be followed through the website http://30dor.stsci.edu.

This paper is organized as follows: we present the motivations for the survey in Section~\ref{starbursts}, and the observing strategy in Section~\ref{obs}. As of the writing of this paper, we have collected half of the data in the near-IR (NIR). The data reduction and the analysis of this portion of the dataset are discussed in Section~\ref{data_analysis}, while the color-magnitude and color-color diagrams (CMDs and CCDs respectively) are presented in Section~\ref{cmds}. A discussion of the results and conclusions can be found in Section~\ref{discussion}.

\section{Tracing the Evolution of a Starburst}
\label{starbursts}

\subsection{The Cosmological Context}
Starbursts are short-lived periods of intense, massive star formation \citep{searle73} that occur in compact regions (10 to 1000 pc) and dominate the overall luminosity of their host galaxy \citep{heckman05}. In their Annual Review
\citet{kennicutt12} reminds us that the star formation rate per unit area in a starburst is much larger than in disk-averaged star formation rate surface densities ($\sim 0.1\, {\rm M_\odot yr^{-1} kpc^{-2}}$). Evidence for recent starburst activity is found in 6\% of the galaxies in the Local Universe \citep[$z\ll1;$][]{lee07, lee09} and in 15\% of those at high redshift \citep[$z>1;$][]{oconnell05, douglas10}, indicating that starbursts are a relatively common phenomenon. The impact of a major starburst on the evolution of a galaxy is dramatic, because it will shape the galaxy dynamics, stellar evolution and chemical compositions in ways that are largely dictated by the intensity and the duration of the bursting event  \citep[i.e.][]{dekel86, maclow99, kennicutt12}. 

Starbursts are studied on two complementary fronts. High-redshift surveys collect information over a range of long look-back times for a large sample of objects, but suffer from limited or no spatial information and crude age diagnostics. Studies in the Local Universe, on the other hand, are affected by the small size of the sample, but can probe discrete structures within the star-forming region using more precise population diagnostics \citep{hunter06, tolstoy09, mcquinn12}. In both approaches the characterization of starburst properties is still rudimentary and sometimes contradictory.

Studies in individual nearby dwarf galaxies and in surveys at redshift higher than $\sim 1.0$ indicate that a starburst may last for 5-10 Myr, suggesting that the violent feedback from the massive stars formed during the burst suppresses for a while any further star formtion \citep{tosi89, ferguson98, annibali03, stinson07, annibali09}. Other studies in the Local Universe, however advocate for much longer star-forming episodes \citep[$> 100\, {\rm Myr;}$][]{konstantopoulos09, mcquinn10}, although it is not clear if their approach allows for sufficiently high temporal resolution. Since the spectral energy distributions (SEDs) of galaxies at high redshift can be interpreted only by assuming a duration for the burst, understanding how and for how long star formation can be sustained during a starburst event has profound implications on our picture for understanding galaxies.  

\subsection{The Tarantula Nebula: A Unique Opportunity to Decipher Starbursts}

The Tarantula Nebula is by far the most luminous and massive star-forming region in the Local Group \citep{kennicutt86}. Covering an area of $\sim 40,000\, {\rm pc^2}$, the Tarantula Nebula is the closest extragalactic giant H{\sc ii} region and is comparable in size to the unresolved luminous H{\sc ii} complexes observed in distant galaxies \citep{oey03, hunt09}. 

With more than 800 spectroscopically confirmed OB stars \citep{evans11}, some of which are among the most massive candidates \citep{crowther10}, 30~Dor is often described as a mini-starburst \citep{leitherer98}. The UV flux coming from 30~Dor is such that it has been proposed as a small scale local analog to the Lyman-Break Galaxies \citep{meurer97, shapley03, heckman04}. Furthermore several authors  \citep[e.g.][]{oconnel78, oconnel95, brandl04} have suggested that the luminous knots observed in bursting galaxies such as M82 are made of (multiple) 30~Dor-like systems. 

By virtue of its location in the LMC \citep[$\sim 50\, {\rm kpc}$;][]{panagia91, pietr13}, with {\it Hubble} we can resolve the Tarantula Nebula into single stars down to the sub-solar mass regime \citep{andersen09, demarchi11}. The low inclination angle \citep[$\sim 30\degr$;][]{nikolaev04} limits the line of sight confusion, and the foreground reddening is low because of the high Galactic latitude. 

The Tarantula Nebula is a very dynamic region both in terms of runaway-star candidates \citep[e.g.][]{evans11} and gas motion \citep{chu}. Since it is a multi-stage star-forming region \citep{walborn97}, where loose associations and very dense star clusters of different ages coexist, the Tarantula Nebula is a noteworthy window in which to study the different modes of star formation and their reciprocal interaction. In particular, observational evidence shows that over the last $\ga 25\, {\rm Myr}$ the star-formation history of  the region (Figure~\ref{f:tarantula}) has been complex :
\begin{itemize}
\item \citet{walborn97} identified multiple generations of stars ranging from $\le0.5\, {\rm Myr}$ to $\sim 25\, {\rm Myr}$ in $1\arcmin - 2\arcmin$ from the core of 30~Dor,  R136.
\item \citet{grebel00} derived an age of $\sim 20 - 25\, {\rm Myr}$ for Hodge~301, a cluster $\sim 3\arcmin$ to the northwest of R136.
\item NICMOS observations at the ``frontier'' between Hodge~301 and R136 showed that this region is strewn with several massive O stars embedded in dense knots of dust, suggesting that stellar feedback from the two clusters may be triggering new episodes of
star formation \citep{brandner01}.
\item $\sim 7\arcmin$ to the west of R136 the super-bubble created by the $\sim 10\, {\rm Myr}$ old OB association LH99 is filled by the expanding supernova remnant N157B, which contains the most energetic pulsar known \citep{chen06}. 
\item Deep HST optical and NIR observations suggest that the core of 30~Dor, R136, is an interacting double cluster \citep{sabbi12}.
\end{itemize}

As these previous studies hint, the Tarantula Nebula region is complicated. While there have been attempts to study in detail some of its specific parts, much of the focus with HST to date has been on its central cluster \citep{walborn99, brandner01, walborn01, andersen09, demarchi11, sabbi12}. These studies have yielded significant results, nevertheless a comprehensive understanding of the Tarantula Nebula in its entirety can only be reached by detailed studies that include other star clusters and surrounding field stellar populations.

\subsection{The HTTP Science Drivers}

\begin{figure}
\epsscale{1.0}
\plotone{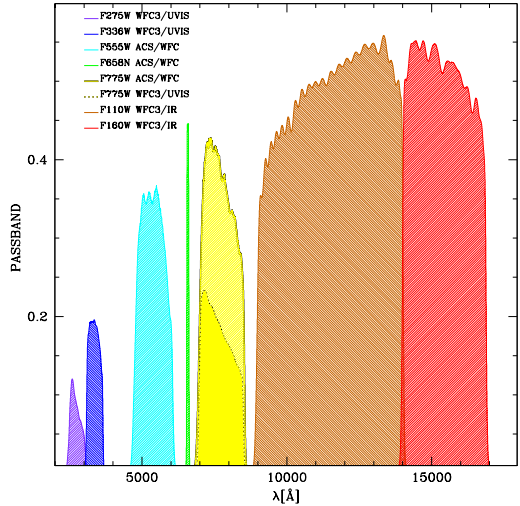}
\caption{\label{f:passband} Passband of the filters used in HTTP as a function of wavelength.}
\end{figure}

The ongoing HST panchromatic survey of the Tarantula Nebula HTTP (PI Sabbi, GO-12939) is built on an existing HST program (PI Lennon, GO-12499) in filter F775W ($\sim {\rm SDSS}\, i$-band). HTTP was awarded 60 orbits of HST time. Figure~\ref{f:passband} shows the wavelength coverage of the survey, once it is combined with the GO-12499 dataset.

The decision to survey the region over such a wide wavelength baseline was driven by several factors. Adding an optical filter to the F775W dataset allows us to reconstruct the star formation history of the entire region and infer the changes in star formation rate with time down to $\sim 0.5\, {\rm M_\odot}$. We preferred the ACS F555W to the wider F606W filter, because the latter, being quite red, would have provided a limited color baseline relative to the F775W filter. Furthermore the H$\alpha$ line falls in the F606W bandpass and in H{\sc ii} regions this can affect the accuracy of the photometry.

\begin{figure*}
\epsscale{1.0}
\plotone{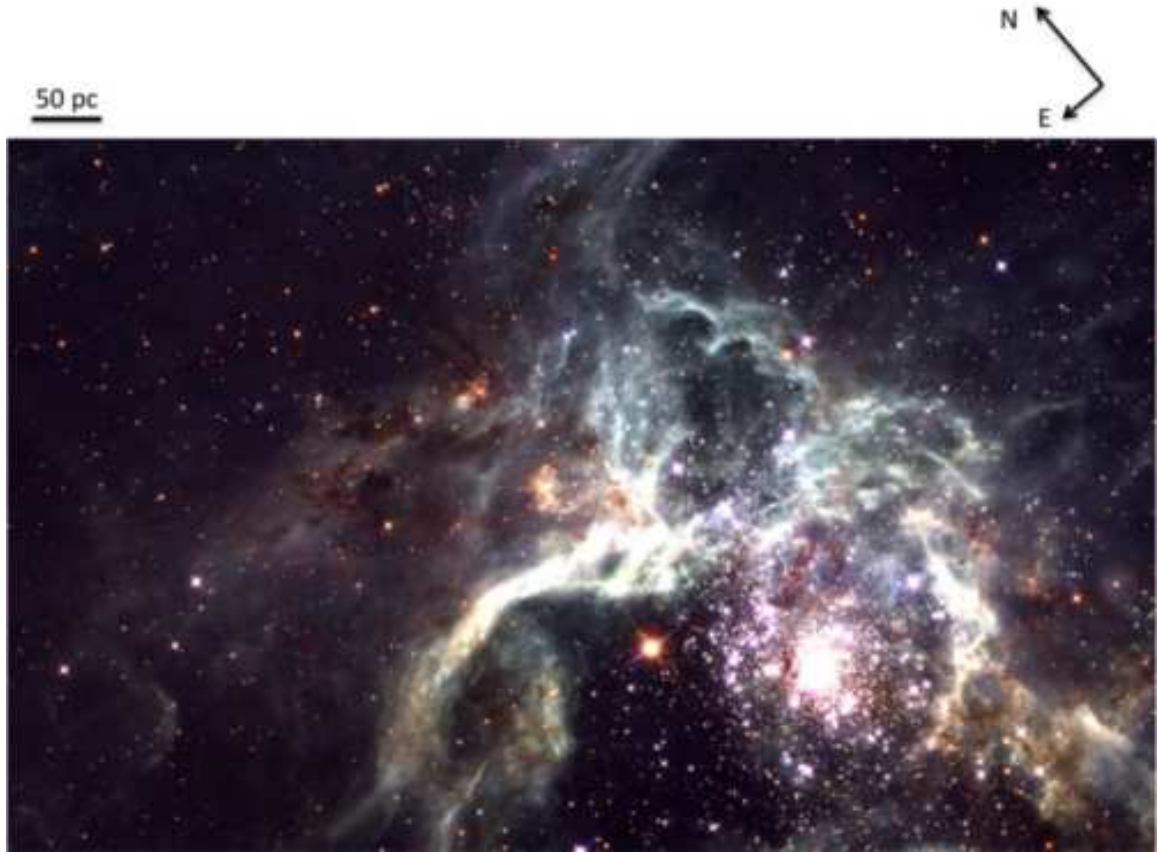}
\caption{\label{f:30dor_ir} HST color-composite image of R136, F775W is in blue, F110W in green and F160W in red. }
\end{figure*}

Supplementing our dataset with two NIR filters allows us to extend our study to the dustiest regions, in order to better constrain the properties of cool stars such as red supergiant, red giant (RGB) and asymptotic giant branch (AGB) stars. Even more important, only NIR observations can identify regions of ongoing star formation that are still embedded in their dusty cocoons (see Figure~\ref{f:30dor_ir}). 

Previous NICMOS observations have shown that star formation took place in the recent past at the gaseous interface between R136 and Hodge~301 \citep{brandner01}. By extending this study to the entire Tarantula Nebula we can identify pockets of ongoing
star formation far from the main clusters and test the constructive properties of stellar feedback predicted by the models \citep[e.g.][]{elmegreen77, vanhala98}. We have selected the F110W and F160W filters, since they provide the highest throughput with the IR channel.

Observations in the near-UV (NUV) have been designed to infer the temperature of the hotter stars and, once combined with optical measurements, constrain the amount of extinction \citep{zaritsky99, romaniello02}. We have chosen the WFC3 filter F336W because of its high throughput and because it is specifically designed to constrain the Balmer break. To better determine the temperature of the hotter stars and to provide the largest baseline possible to characterize the extinction law we added the even bluer filter F275W.

Finally we are observing the Tarantula Nebula with the narrow-band ACS filter F658N (corresponding to H$\alpha$). Young ($<5\, {\rm Myr}$) low-mass pre-main sequence (PMS) stars can be easily identified in the optical CMD as a population of faint red objects, and can be used to trace the extent of young star-forming regions \citep{sabbi07, cignoni2010, gouliermis2012, sabbi12}. With the aging of a stellar population, however, the colors of the PMS stars become too blue to be distinguished from the low-mass MS stars of the field, an effect that becomes even more severe when extinction is highly variable, as is the case for the Tarantula Nebula. In previous studies \citep{panagia00, demarchi2010, demarchi11} we have shown that a combination of broad- and narrow-band filters can be used to identify low-mass PMS stars with active mass accretion, and that this approach is very efficient in picking up older ($>10\, {\rm Myr}$) and more diffuse episodes of star formation. In addition, the same strategy will allow us to identify main-sequence and evolved stars with H$\alpha$ excess, such as Be and B[e] stars \citep{grebel97}.

In summary the high sensitivity, spatial resolution, and broad wavelength coverage of HTTP will allow us to reconstruct the Tarantula's star-formation history in space and time on a parsec scale. In particular we will be able to (1) characterize  when and where star formation occurred; (2) establish the length and the strength of the star-forming episodes and their spatial scale;  (3) depict the life cycle of star clusters; (4) portray the role of stellar feedback in shaping the evolution of a starburst; and (5) probe the universality of the stellar initial mass function. In addition, a Bayesian fitting of the stellar SEDs will allow us to accurately determine R$_V$, and therefore supply information about the size of the dust grains and their spatial distribution (K. Gordon et al. 2013, in preparation).

\section{Description of the Observations}
\label{obs}

In both datasets described in this paper (GO-12499, PI Lennon; and GO-12939, PI Sabbi) the Wide Field Channel (WFC) of ACS is used in parallel with either the UVIS or the IR channels of WFC3 to maximize the efficiency of the observations. Since there are no filters in common between the WFC3/IR channel and the WFC3/UVIS or ACS/WFC, in this paper we will simply refer to the instrument (ACS and WFC3), without specifying the channel. 

\begin{figure}
\epsscale{1.0}
\plotone{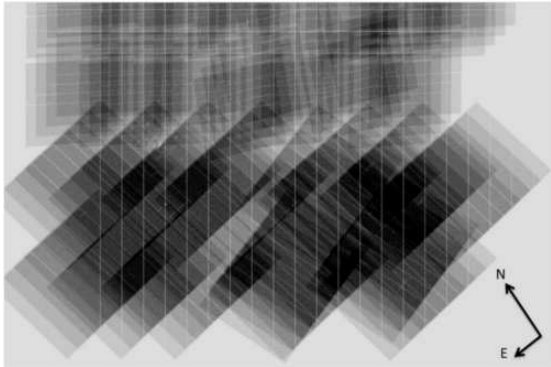}
\caption{\label{f:depth} Map of the depth of the mosaic obtained by stacking all the deep exposures of the GO-12499 dataset. The darker the image, the larger the number of overlapping images. WFC3 is up and ACS is down. The image is rotated by 90$\degr$ compared to Figure~\ref{f:tarantula} .}
\end{figure}

\subsection{The GO-12499 Dataset}

The {\it HST} program GO-12499 (P.I. Lennon) represents the first epoch of a proper-motion survey of massive stars in the 30~Doradus region. The survey covers a projected area of $\sim 14\arcmin \times 12\arcmin$, corresponding to $\sim 210\times180\, {\rm pc}$ at the distance of the LMC.

A total of 15 pointings were used to map the whole region. All the observations were taken between  3 October and 29 October 2011, using WFC3 as primary instrument and ACS in parallel. The orientation angle of the mosaic was chosen to include the very massive runaway VFTS\#16 \citep{evans10}.

Since the goal of the program is to determine the positions of the stars with exquisite accuracy, all the observations were acquired in a single filter. Filter F775W was chosen in an effort to minimize the effect of the strong and variable nebular emission that pervades the entire region.

Each pointing was observed for one orbit of HST tim, and each orbit consisted of one short and four deeper exposures with each camera. The exposure times were dictated by the buffer dumping of WFC3 and by the length of the orbit. To give the survey as much spatial uniformity as possible, observations were stepped in such a way that no star would fall in the inter-chip gap in more than one of the WFC3 exposures. 

The full dataset consists of 15 exposures of 35 sec, 14 of 500 sec and 45 of 700 sec with UVIS, and 15 of 32 sec, 14 of 377 sec and 45 of 640 sec with ACS for a total of 148 exposures. Once stacked together the deep exposures result in a mosaic of $\sim28,000\times 22,000$ pixels, with a $40.00\, {\rm mas\, pixel^{-1}}$ scale. Figure~\ref{f:tarantula} shows the mosaic, with colors added based on ground based data.  Figure~\ref{f:depth} shows the depth map of the deep stacked image.

The CCD detectors on board HST are subject to a flux of energetic particles that progressively damage the silicon lattice of the detectors and create charge traps that redistribute the flux from one pixel to the other during the detector readout process. As a result of this cumulative radiation damage, the charge transfer efficiency (CTE) of the CCDs on board HST is progressively degrading. We used the algorithm described by \citet{anderson10} to correct the effects of the degrading CTE directly on the images. This algorithm is automatically applied by the ACS calibration pipeline. The routine we used to correct WFC3 data for CTE can be downloaded from the WFC3 website\footnote{http://www.stsci.edu/hst/wfc3/tools/cte\_tools}.

\subsection{The GO-12939 Dataset}

\begin{figure*}
\begin{centering}
\includegraphics[scale=0.25]{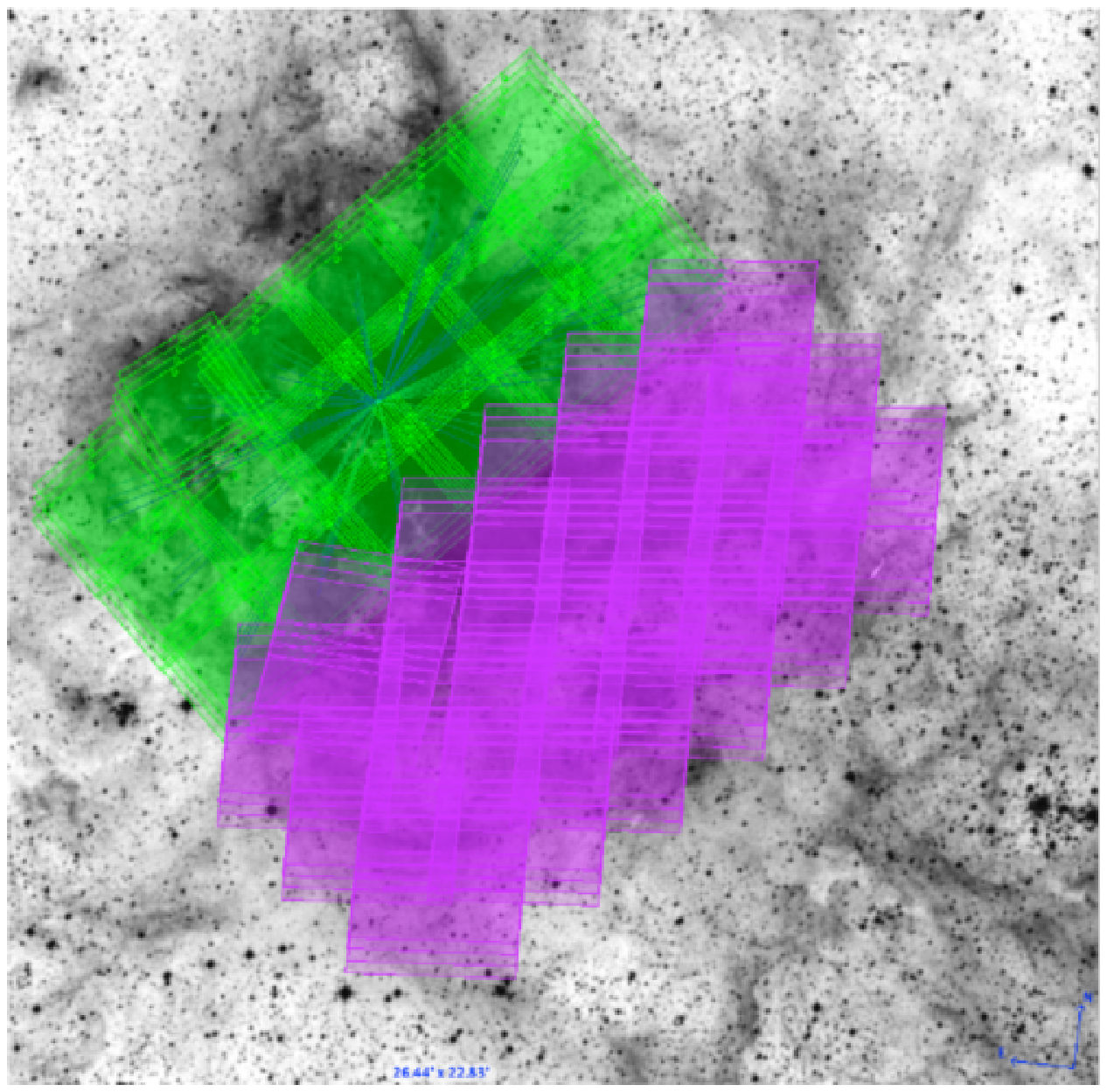}\includegraphics[scale=0.25]{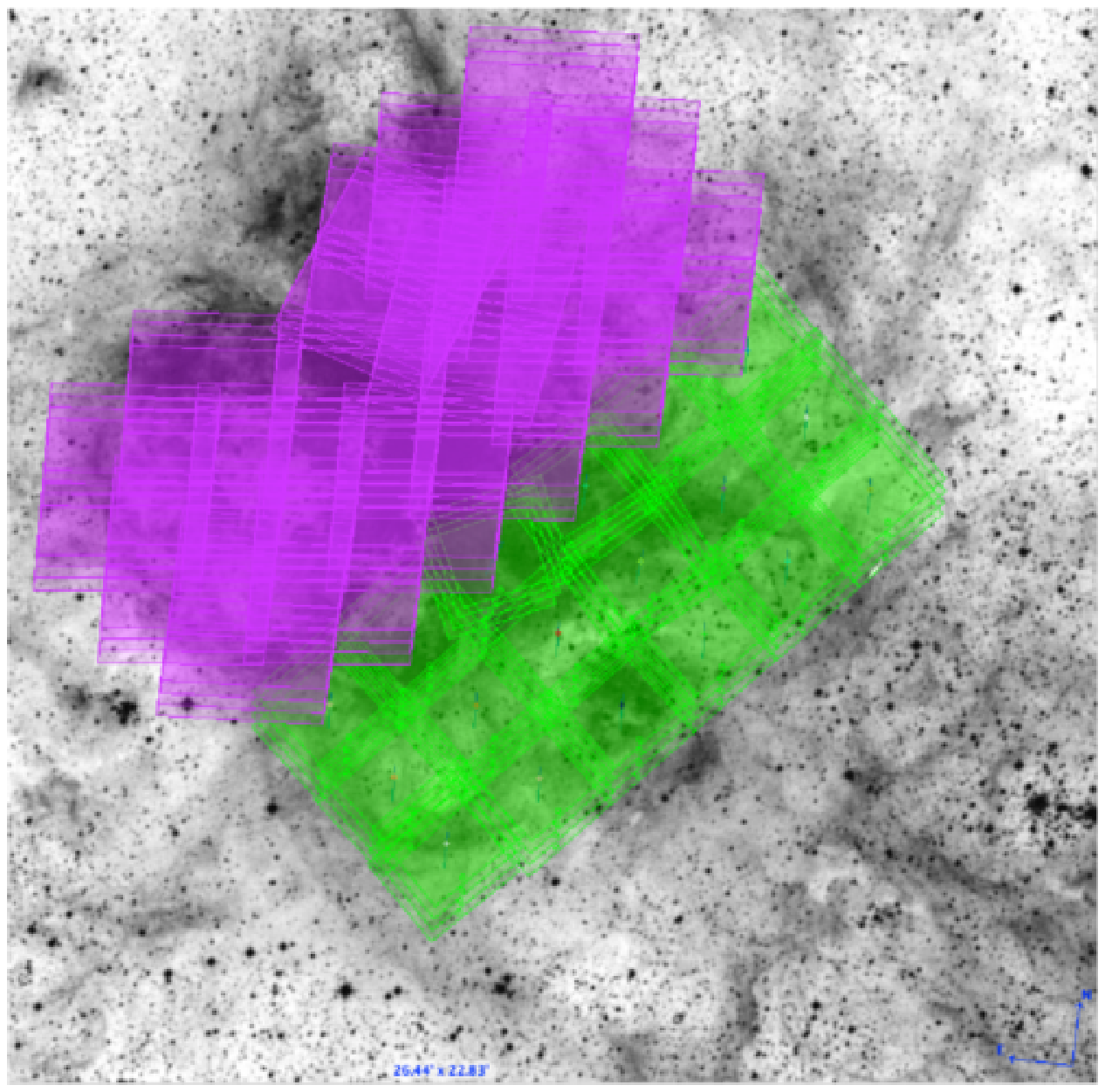}
\par\end{centering}
\begin{centering}
\includegraphics[scale=0.25]{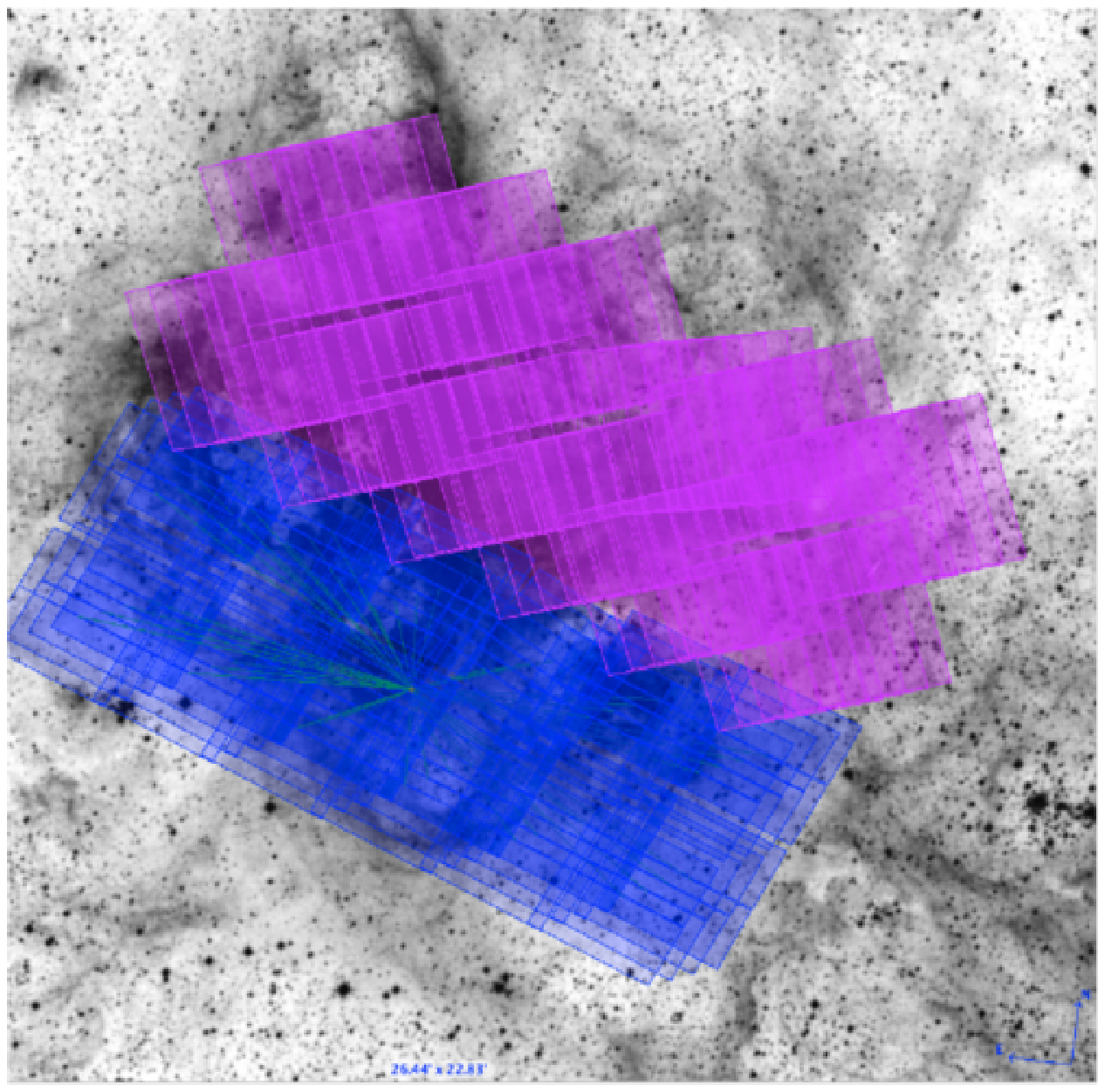}\includegraphics[scale=0.25]{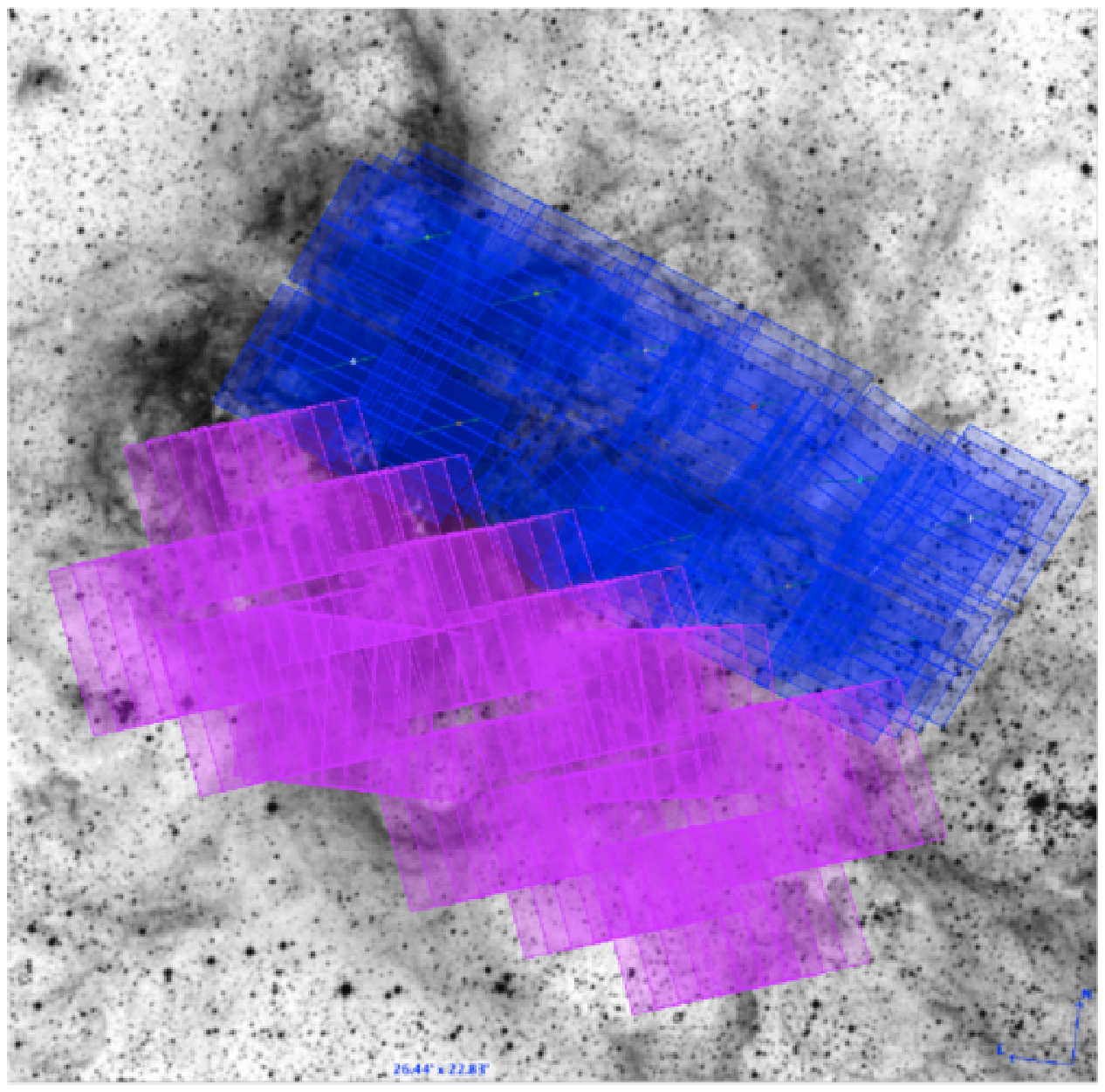}
\par\end{centering}
\caption{\label{f:pointings} {\it Upper Panels: } Positions of the HST pointings in the filters F110W, F160W (WFC3) and F658N (ACS), superimposed on the $33\arcmin\times21\arcmin$ DSS image of 30 Doradus. WFC3 is in green and ACS in magenta. {\it Lower panels:}  Positions of the HST WFC3 (F275W and F336W filters) and ACS (filter F555W) pointings.  WFC3 is in blue and ACS is in magenta. Figure~\ref{f:tarantula} is rotated by 45$\degr$ with the respect to these plots.}
\end{figure*}

As was done for program GO-12499, we are observing the field with WFC3 as primary instrument and ACS in parallel. Visits are organized in four blocks of visits. Figure~\ref{f:pointings} shows the 36 NIR and 24 UVIS WFC3 pointings superimposed on the DSS image of 30~Doradus, together with the ACS parallels. Each pointing corresponds to a single orbit of HST time and consists of a four-step dither-pattern.

The first and third blocks are observed with WFC3/IR and ACS. During each orbit we acquire two exposures in the WFC3 F110W and 2 in the F160W filters, while ACS is used to collect four observations in the F658N filter (In Figure~\ref{f:pointings}, upper panels, WFC3  is in green and ACS is in magenta). 

In the second and fourth blocks, each orbit begins with the acquisition of a short exposure in the F336W filter with WFC3 and a short ACS image in the F555W filter. We then collect two images in the F336W and two in the F275W filters. At the same time, we acquire four exposures in the F555W filter with ACS (in Figure~\ref{f:pointings}, lower panels, WFC3 is in blue and ACS is in magenta).

Half of the IR observations were acquired between 12 and 19 December 2012 and are described in the next section. The remaining half is scheduled for the summer of 2013. The NUV observations will be collected in April and in September 2013. This observing strategy is designed to exploit the natural rotation of the telescope during the year. 

The filters, the number of exposures, and the exposure times used in HTTP are summarized in Table\ref{t:obs}. Updates on the status of the survey can be found at http://30dor.stsci.edu.
 
Observations with very low background ($<10\, e^-$ for WFC3 and $<50\, e^-$ for ACS, such as those acquired in the NUV and/or through narrow-band filters) suffer large losses for very faint sources, to the point that faint sources can completely vanish during the readout transfers. Since Cycle 20, it has been possible to mitigate the effects of the degrading CTE by adding an extra amount of photons through an internal illumination. This process, called ``post-flash'', has the effect of artificially increasing the background level.  All the NUV observations of HTTP are acquired using ``post-flash''. We decided to not use post-flash for the H$\alpha$ exposures, since we estimated that the background would be sufficiently high. Even with the background increased, all sources still suffer some CTE loss, and it is necessary to make a photometric correction for the detected sources. For this reason we will correct all the data acquired in the NUV and visual regimes using the Anderson \& Bedin algorithm.
 
\begin{table*}
\begin{center}
\caption{{Journal of observations}\label{t:obs}}
\begin{tabular}{ccccc}
\tableline
\tableline
{\bf Instrument} & {\bf Filter Name} & {\bf Number of Exposures} & {\bf Exposure Time} & {\bf Post-Flash}\\
\tableline
WFC3/UVIS & F275W & 1 & 467  & YES\\
WFC3/UVIS & F275W & 1 & 697  & YES\\ 
WFC3/UVIS & F336W & 1 &   14  &  YES\\
WFC3/UVIS & F336W & 2 & 700 & YES\\ 
ACS/WFC & F555W & 1 &   13  &  NO\\ 
ACS/WFC & F555W & 1 & 337  &  NO\\
ACS/WFC & F555W & 3 & 640  &  NO\\
ACS/WFC & F658N & 1 & 300  &  NO \\
ACS/WFC & F658N & 3 & 640 &  NO\\
ACS/WFC & F775W & 1 & 32  & NO \\
ACS/WFC & F775W & 1 & 377  & NO \\
ACS/WFC & F775W & 4 & 640  & NO\\
WFC3/UVIS & F775W & 1 & 35  & NO \\
WFC3/UVIS & F775W & 1 & 507  & NO \\
WFC3/UVIS & F775W & 4 & 690  & NO \\
WFC3/IR & F110W & 1 & 300 & - \\
WFC3/IR & F110W & 1 & 640 & - \\
WFC3/IR & F160W & 2 & 640 & - \\
\tableline
\end{tabular}
\end{center}
\end{table*}

\section{Data Analysis }
\label{data_analysis}

The analysis of the F775W (GO-12499) and  the NIR data was carried out directly on the bias-subtracted and flat-fielded exposures processed by the standard HST calibration pipelines CALACS and CALWF3. The outputs from the HST calibration pipelines are still in the raw-detector pixel frame and thus retain several kinds of spatial distortions that have to be taken into account. 

\subsection{Preparation of the Reference Frame}
\label{reference_frame}

Observations in the F775W filters cover a projected area of $\sim 14\arcmin \times 12\arcmin$. The first step in reducing the data was to create a distortion-free reference frame and relate the astrometry and photometry of each exposure to this frame. To achieve this, in each F775W deep exposure we measured positions and fluxes for all the sources that had more than 100 counts in each exposure and no brighter neighbors within a 5 pixels radius using the program {\tt img2xym\_WFC.09x10} \citep{anderson06}\footnote{The program was originally designed for ACS/WFC data, but  versions for both the WFC3 channels are now available.}. This program uses libraries of empirical PSFs to account for the spatial variations caused by the optics of the telescope and the variable charge diffusion in the CCD and creates a geometric-distortion corrected list of sources. 

We used the Two Micron All Sky Survey (2MASS) catalog \citep{skrutskie06} as an initial astrometric reference frame. We then found common stars between each image list and 2MASS to define the general six-parameter linear transformation between the distortion-corrected frame of each exposure into the 2MASS-defined frame. We used these transformations to collate the individual star lists in the reference frame and determine the average position for $\sim 110,000$ bright, isolated, unsaturated stars that in the deep exposures have three or more coincident detections. 

For the ACS data we initially used the geometric distortion and PSF library of \citet{anderson06}. Both the PSF and distortion solution were determined using data acquired before the last Hubble servicing mission (SM4). A careful inspection of the position residuals showed that the PSF has developed a 2\% asymmetry and the distortion solution has changed by $\sim 4\%$ of a pixel since SM4. Although this is a very small change, it is sufficient to introduce noticeable errors in very wide mosaics, therefore we used all the bright and isolated stars found in the ACS field of view to derive a new PSF and update the geometric distortion. Since all the observations in this program were acquired with very similar rotation angle, we were not able to independently solve for the ACS skew term, however we used the common stars between ACS and WFC3 to calibrate it relative to WFC3. 

Finally we improved the internal quality of the reference frame by iteration. The r.m.s.\ residuals of the average  positions were less than 0.01 pixel in each coordinate. In this process we have mapped the ACS pixel scale to the size of the WFC3 UVIS channel, so that in all the catalogs a pixel corresponds to $0.040\arcsec$. The final reference list was then used to compute the final astrometric transformation of each exposure (for both ACS and WFC3) into the reference frame.

\subsection{The Source-Finding Routine}

\begin{figure}
\epsscale{1.0}
\plotone{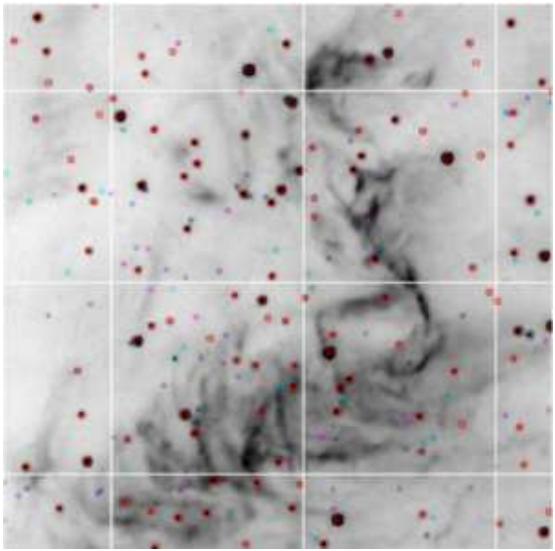}
\caption{\label{f:no_rusco} Stars found in a $360\times 360$ pixel region in the F775W filter. Sharp filaments and unresolved knots of gas are visible. Different colors and sizes of the circles correspond to different finding iterations in the photometry. }
\end{figure}

The stars used in Section~\ref{reference_frame} to align the single exposures to the reference frame were found in a single pass through each image. This approach will find almost all the bright stars in a field, but will miss many of the obvious faint sources, especially those in the wings of bright objects. 

To identify the fainter sources and push the photometry as deep as possible, we used an evolution of the program described in \citet{andreson08}. The new program will be fully described in a forthcoming paper (J. Anderson et al., in preparation). 

The previous version of the code was designed to work only with ACS and could handle only two filters.  This new routine can handle multiple instruments and several filters at the same time.  The routine first finds the bright stars, then subtracts them and searches for fainters stars in the subtracted images.  It performs several such finding iterations (Figure~\ref{f:no_rusco}), using detection thresholds that depend on the number of exposures available at each point in the field. This approach is particularly useful when, as in the case of our survey, the coverage is not uniform (in some regions we have only one or two exposures, while in others we have up to 14 overlapping exposures, Figure~\ref{f:depth}).  An inspection by eye shows that the code efficiently avoids spurious detections caused by the diffraction spikes of saturated stars and sharp gaseous filaments (Figure~\ref{f:no_rusco}).

We were able to identify more than $1.1\times 10^6$ sources in the F775W observations. The photometry was calibrated to the VEGAmag photometric system using the Zero Points listed on STScI webpages\footnote{http://www.stsci.edu/hst/wfc3/phot\_zp\_lbn and http://www.stsci.edu/hst/acs/analysis/zeropoints}. These Zero Points are derived for images combined using ``astrodrizzle'', thus they cannot be directly applied to our {\tt flt}-based photometric catalogs. To calibrate the  {\tt flt}-based catalogs into the Vegamag system we selected several isolated bright stars in the drizzled images and measured their magnitudes using 0.5\arcsec aperture photometry for ACS and 0.4\arcsec for WFC3 data and applied the appropriate Zero Point. For these stars we then determined the difference between the {\tt flt}-based magnitude and the magnitude in the Vegamag system and used the average differences as the final Zero Points. 

The F775W photometric catalog was used to measure the sources in the IR images. Of the initial $1.1\times 10^6$ sources, $\sim 480,000$ are in the region observed with the first block of IR data.  

\begin{figure}
\epsscale{1.0}
\plotone{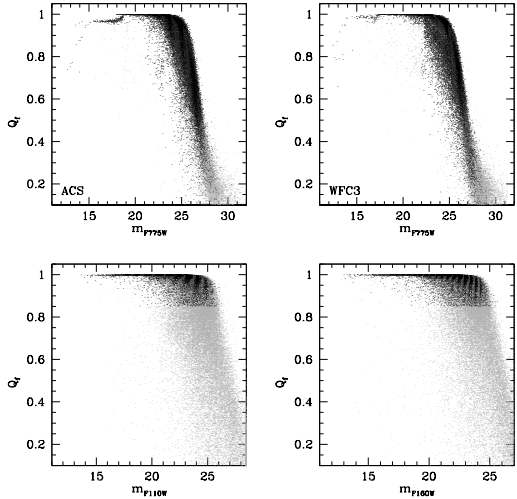}
\caption{\label{f:horror} From the top to the bottom, clockwise, {\tt Q} parameters as a function of magnitude for the filters F775W (ACS and WFC3 respectively), F160W and F110W. Only 10\% of all the sources detected in the F775W filers and 20\% of the sources found in the F110 and F160W filters are shown. Objects that meet the selection criteria are in black.} 
\end{figure}

In addition to finding and measuring stars, the reduction routine also produces several diagnostics to help identify the stars that are well-measured.  One such parameter is {\tt Q}, which reports the linear-correlation coefficient between the PSF and the stellar image:  {\tt Q}$ =1.0$ is a perfect fit.  This parameter is particularly useful in identifying unresolved objects such as background galaxies and blended sources. We also used the r.m.s.\ of the individual photometric measurements about the mean to help identify well measured stars.  Figure~\ref{f:horror} shows {\tt Q} as a function of magnitude for the two F775W filters (ACS and WFC3 are showed separately) and for the NIR data.  

In both ACS and WFC3 images acquired with the F775W filters, stars brighter than $m_{\rm F775W}=15$ are saturated even in the shorter exposures. As a result the {\tt Q} parameter for these stars differs significantly from 1. We used aperture photometry to recover the flux of these sources, and we did not apply any selection criteria. For ACS we then selected the sources in the magnitude range between $15\le m_{\rm F775W_{ACS}} <21$ with {\tt Q}$>0.95$, while for sources fainter than $m_{\rm F775W_{ACS}}\ge21$ we selected only those objects with {\tt Q}$>0.92$ or photometric error $<0.5$. Similarly, for the F775W  WFC3 filter in the magnitude range between $15\le m_{\rm F775W_{WFC3}} <22$ we selected only the objects with  {\tt Q}$>0.95$ and for fainter sources the photometric error had to be better then 0.5. 

Because of the coarser resolution of the IR channel, crowding is more severe in the filters F110W and F160W and the parameter {\tt Q} does not provide an estimate of the quality of the photometry that is as good as in the case of the optical wavelengths. Moreover for most of the field in the NIR we have only two measurements per filter, therefore to estimate the photometric error we used the signal to noise instead of the r.m.s.\ of the individual photometric measurements. As a result for both the F110W and F160W filters we selected only those objects with {\tt Q}$>0.85$ and photometric error smaller than 0.1.  After these selections the combined ACS$+$WFC3 catalog in the F775W filter contains more than 366,000 stars, of which $\sim150,000$ that are detected in both the F110W and F160W filters.

\section {Color-Magnitude and Color-Color Diagrams}
\label{cmds}

\begin{figure*}
\epsscale{1.0}
\plotone{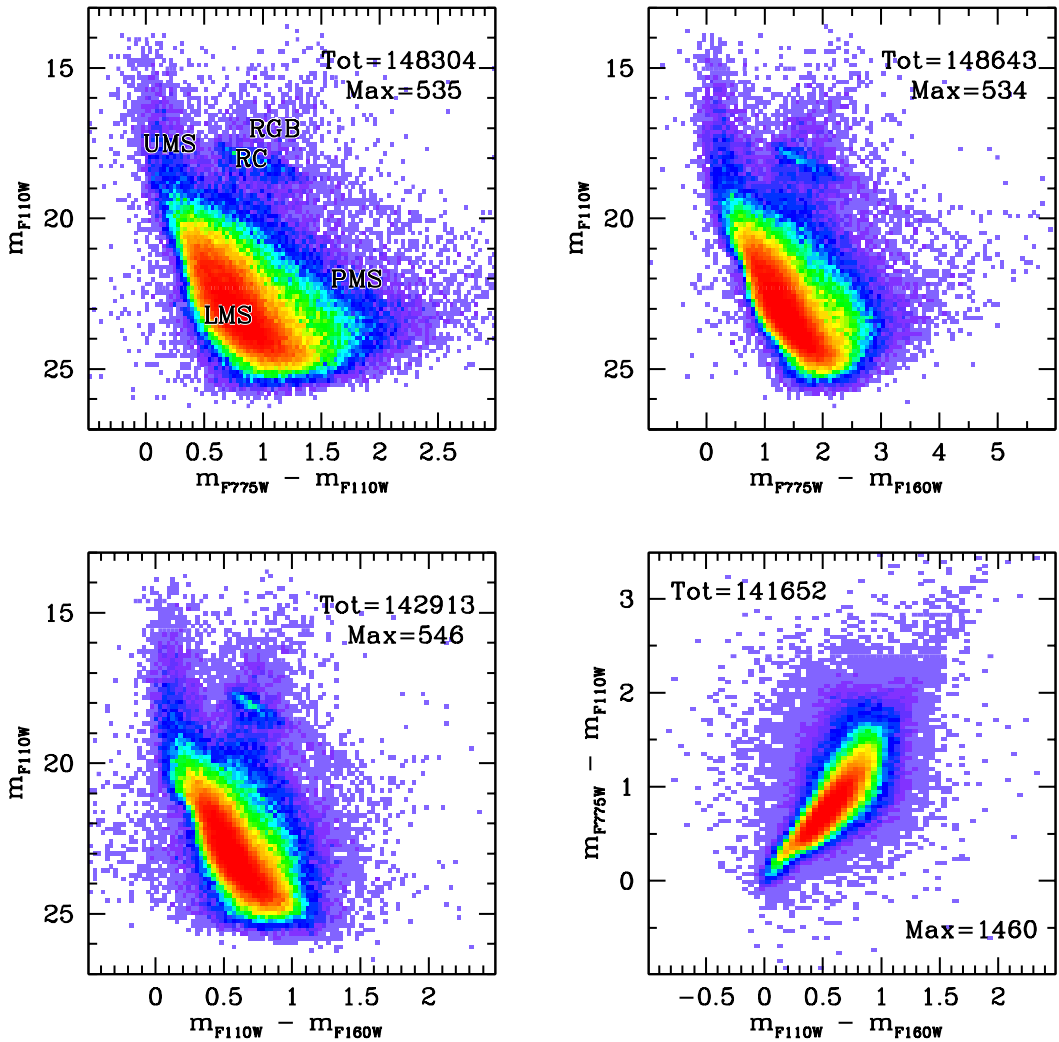}
\caption{\label{f:hess} From the top to the bottom, clockwise, CMDs in the filters $m_{\rm F110W}$ vs $m_{\rm F775W}-m_{\rm F110W}$,  $m_{\rm F110W}$ vs $m_{\rm F775W}-m_{\rm F160W}$, $m_{\rm F110W}$ vs $m_{\rm F110W}-m_{\rm F160W}$, and CCD  $m_{\rm F775W}-m_{\rm F110W}$ vs $m_{\rm F110W}-m_{\rm F160W}$. In each diagram both the total number and the maximum number of stars per bin of magnitude and color are reported. The most important evolutionary sequences are highlighted in the upper-left panel for reference. }
\end{figure*}

Figure~\ref{f:hess} shows the Hess CMDs and CCD for different combinations of the F775W, F110W, and F160W filters. The main evolutionary features are highlighted for reference in the top left-panel. In this CMD stars brighter than $m_{\rm F110W}\sim 20$ and bluer than $m_{\rm F775W}-m_{\rm F110W} \sim 0.5$ are in the the upper main sequence (UMS). These are intermediate- and high-mass stars. The UMS is representative of a stellar population younger than $\sim 1\, {\rm Gyr}$. 

Stars fainter than $m_{\rm F110W} \sim 22$ and bluer than $m_{\rm F775W}-m_{\rm F110W} \sim 1$ are in the lower main sequence (LMS). This is a mixture of stars with ages spanning the age range from potentially very young to several Gyr old.

Stars brighter than $m_{\rm F110W}\sim 21$ and redder than $m_{\rm F775W}-m_{\rm F110W} \sim 0.5$ are in the sub giant branch (SGB), red giant branch (RGB) and red clump (RC) evolutionary phases. These sources are characteristic of an evolved old ($\ga 1-2\, {\rm Gyr}$) stellar population. In all the CMDs, the RC is elongated, extending over more than $\sim 1.1$ magnitudes in the F110W filter. Similarly both the UMS and the RGB are broad suggesting that the reddening is variable and differential extinction often is significant across the field. 

The lower ($m_{\rm F110W} \ga 21$) right ($m_{\rm F775W}-m_{\rm F110W} \ga 1$) corner of the CMD is usually populated by young, intermediate- and low-mass PMS stars. These sources have not yet started the hydrogen burning in their core and are still contracting toward the Zero Age Main Sequence. 

In other young ($\la 5\, {\rm Myr}$ old) clusters in the MCs, the PMS stars are normally well separated from the LMS stars in the CMDs derived from HST observations. This is not the case for the CMDs shown in Figure~\ref{f:hess} not only because differential reddening shifts LMS stars toward redder colors, but also because with the aging of the stellar population PMS stars move towards bluer colors. As the presence of the $\sim 20 - 25\, {\rm Myr}$ old cluster Hodge~301 indicates, star formation in the Tarantula Nebula region has been active for several Myr, therefore in the CMDs of the entire field of view, LMS and PMS are overlapping. 

\subsection{Spatial distribution of the stellar populations}

\begin{figure*}
\epsscale{1.}
\plotone{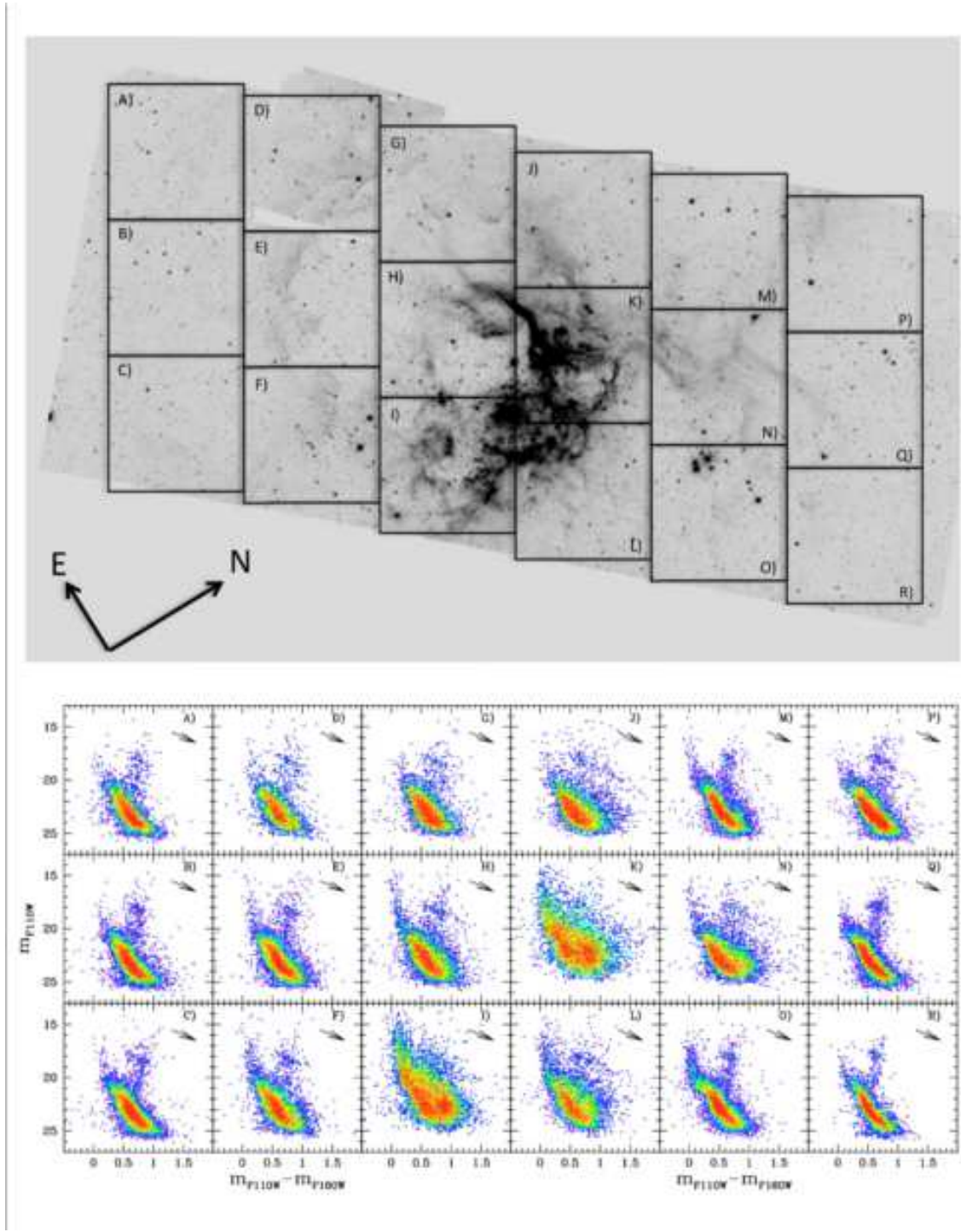}
\caption{\label{f:cmdsa} {\it Upper Panel} Mosaic of the images acquired in the F160W filter. The mosaic has been divided into 18 regions of $\sim 3000\times 3000$ pixels. The corresponding 18 $m_{\rm F110W}$ vs $m_{\rm F110W}-m_{\rm F160W}$ CMDs are shown in the {\it Lower Panel}.}
\end{figure*}

To better understand the spatial distribution of the identified stellar populations, and at the same time, highlight how the reddening changes as a function of position, we have divided our catalog into 18 regions of $3,000\times 3,000\, {\rm pixel}$ each, corresponding to $\sim 29\, {\rm pc} \times 29\, {\rm pc}$. Figure~\ref{f:cmdsa} shows the location of these regions on the F160W mosaic ({\it Upper Panel}) and the corresponding 18 $m_{\rm F110W}$ vs $m_{\rm F110W}-m_{\rm F160W}$ CMDs ({\it Lower Panel}). 

The inspection of these CMDs shows that the UMS is always well populated, with the exception of regions A, B, D, G, Q,  and R. These 6 regions are likely sampling the field of the LMC. 

The majority of R136 is in region I, and the NE clump identified by Sabbi et al (2012) is in region K. The CMDs for both these regions show a large number of red and faint PMS stars, as well as very well populated UMSs. Both R136 and the NE clump are very young systems ($\le 5\, {\rm Myr}$), and the PMS stars are well separated from the LMS stars. PMS stars are likely also present in regions L and N. 

Hodge~301 is in region O. The UMS of the corresponding CMD is well populated in agreement with the young age of the cluster. At these ages PMS stars are too blue to be distinguished from the LMS stars using broad-band photometry only.
 
The RC and the RGB vary from relatively compact and well defined (as in regions B, E, M, O, Q, and R) to extremely broad. For example in region J, the RC covers more than 4 magnitudes in the F110W filter, a typical signature of differential reddening. In their analysis of the reddening in the LMC, bsed on the RC, \citet{haschke11} find that not only 30~Dor exhibits the highest reddening of any region in this galaxy, but also the highest differential reddening. 

It is interesting to note that young and old stars are affected by different amounts of extinction, with the reddening being on average less severe for the young stars, than the old ones. This effect is particularly evident for example in region H, where the UMS is narrow and well defined, but the magnitude of the RC ranges between $18\la m_{\rm F110W} \la 20$. This is apparently at variance with conclusions by \citet{zaritsky99}, who found that on average in the LMC old stars are less obscured by dust than hot young stars. 

In deriving the reddening map \citet{zaritsky99} used only those stars that were detected at the same time in the 4 bands U, B, V and I. This selection criterion penalizes the most extinguished evolved stars, therefore the average amount of reddening derived for the RGB stars may have been underestimated. As shown in \citet{haschke11} in these highly extinguished regions the distribution of the reddening values is not Gaussian, but shows an extended tail towards the higher values. 

It has to be noted that  most of the UMS stars found in the Tarantula Nebula region are associated with the starburst that originated R136 and can be all considered to be at the same distance from us. The RGB stars, on the other hand, are evenly distributed in the entire body of the LMC and are more affected by the 3D structure of the galaxy. If for example 30 Dor is on the nearer side of the LMC disk, the majority of the RGB stars in our field would be beyond 30 Dor and will be extinguished by the dust associated to the Tarantula Nebula. We will be able to better characterize this issue when we observe the region in bluer filters (F275W, F336W, and F555W).  

\begin{figure*}
\epsscale{1.0}
\plotone{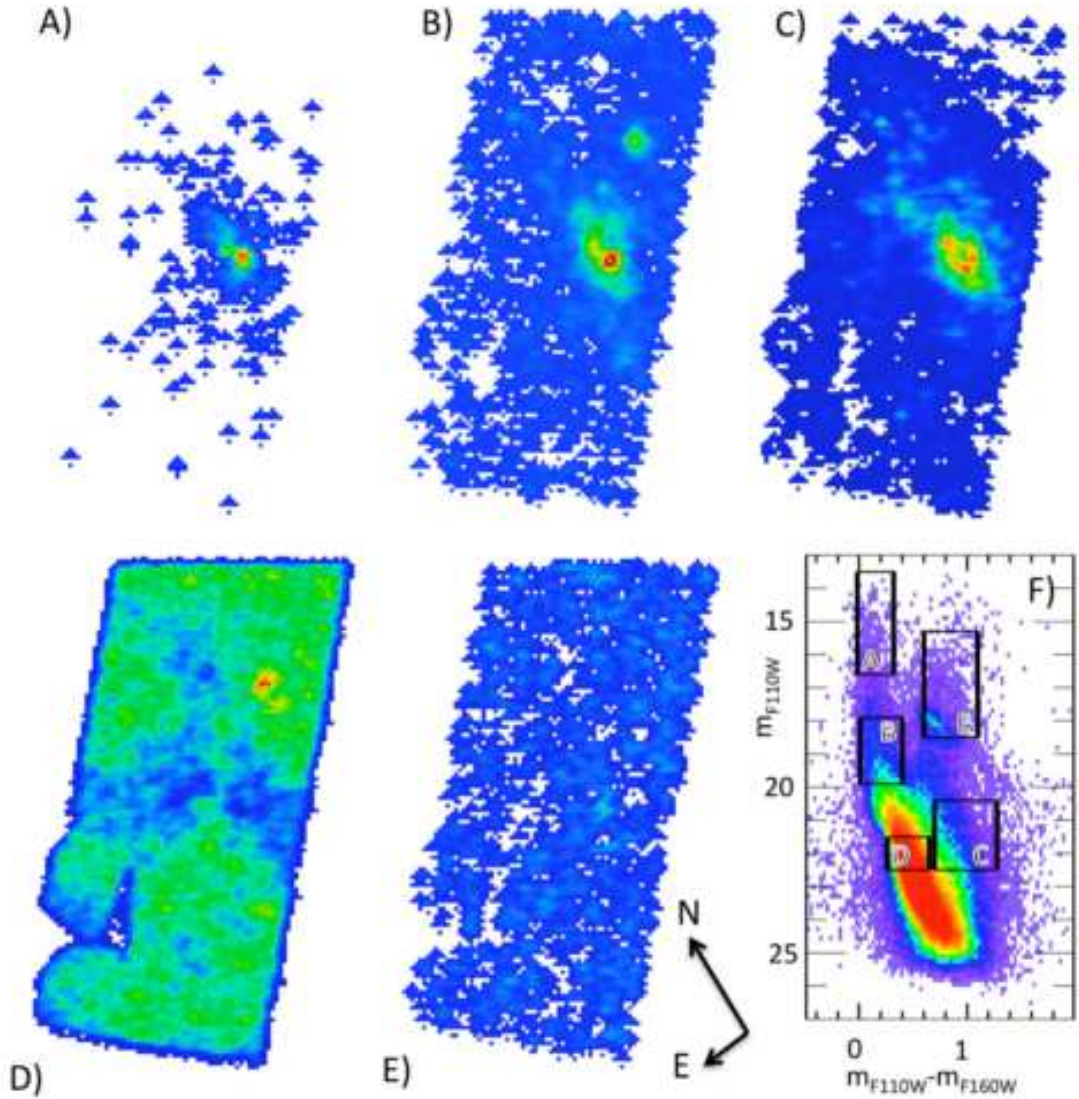}
\caption{\label{f:mappe} Spatial distribution of the massive UMS stars (panel A), intermediate mass stars in the UMS (panel B), PMS stars (panel C), LMS stars (panel D) and RGB stars (panel E). The regions of the $m_{\rm F110W}$ vs $m_{\rm F110W}-m_{\rm F160W}$ CMD used to select the stellar populations are shown in panel F.}
\end{figure*}

The CMDs in Figure~\ref{f:cmdsa} suggest that stellar populations of different ages are not uniformly distributed. To better highlight this effect, in Figure~\ref{f:mappe} we have compared the spatial distribution of stars selected from different evolutionary sequences. 

In panel A we have selected UMS stars brighter than $m_{\rm F110W}< 16.6$, and bluer than $m_{\rm 110W}-m_{\rm 160W}<0.3$. These stars are younger than $\sim 20\, {\rm Myr}$ and are mostly concentrated around R136 and the NE clump. 

Panel B shows the distribution of stars in the magnitude range between $17.9 < m_{\rm F110W}< 19.9$, and bluer than $m_{\rm F110W}-m_{\rm F160W}<0.4$. At the distance of the LMC this correspond to a mass range between $\sim 3$ and $\sim 7\, {\rm M_\odot}$. After $\sim 2\, {\rm Myr}$ the most massive PMS stars in this mass range are already merging into the UMS, while a $\sim 3\, {\rm M_\odot}$ star will still be on the MS after $\sim 150\, {\rm Myr}$. Therefore panel B shows the distribution of stars in the age range between $\sim 2 - 150\, {\rm Myr}$. R136 and the NE clump are still very well defined. The cluster to the North-West of the image is Hodge~301. A careful inspection of panel A shows that Hodge~301 is also visible in that map. 

Sources in panel C have colors between $0.7< m_{\rm F110W}-m_{\rm F160W}<1.3$ and magnitudes between $20.5 < m_{\rm F110W}< 22.5$. This is the locus of $\la 1\, {\rm M_\odot}$ PMS stars. These sources are likely younger than $\sim 5\, {\rm Myr}$. Hodge~301 is not visible in this region, while a ``chain'' of young clustered objects connects the NE clump to the northern corner of the map. Some of these clusters coincide with the embedded massive O stars found in the dense knots of dust at the interface between R136 and Hodge~301, already identified by \citet{brandner01}. Others are aligned with one side of the shell of ionized gas that surrounds a soft and bright X-ray emitting bubble \citep{meaburn84, wang91, townsley06}. Although it is possible that the feedback from the two massive clusters NGC~2070  and Hodge~301 may have triggered the star formation in the nearer clumps, the majority of the clusterings are at a projected distance from R136 between 30 and 60 pc, too far away to be affected by R136 in the less then $\sim 2\, {\rm Myr}$. Alternatively these systems may result from the collapse of residual pockets of gas formed during the fragmentation of the giant molecular cloud that formed R136 \citep[i.e.][and references therein]{bonnell11}. 

LMS stars (panel D) have been selected in the color range  $0.26< m_{\rm F110W}-m_{\rm F160W}<0.65$ and magnitudes between $21.5 < m_{\rm F110W}< 22.5$. These sources have ages between a few tens of Myr and several Gyr. The spatial distribution of these sources is quite uniform in agreement with our assumption that we are looking at the stars in the field of the LMC. The only visible cluster in panel D is Hodge~301, in agreement with the fact that old PMS stars cannot be separated from LMS stars using broad-band photometry only. 

The number of LMS stars is significantly lower in the regions that in panel C correspond to sites of very recent star formation. This could be a further indication that 30 Dor is on the nearer side of the LMC disk. The spatial distribution of RGB stars (panel E, $0.6< m_{\rm F110W}-m_{\rm F160W}<1.1$, $15.3 < m_{\rm F110W}<18.5$) is noisy because of low number statistics, but on average appears quite uniform. 

\section{Discussion and Conclusions}
\label{discussion}

We have discussed the observing strategy and presented preliminary results from the first half of the IR observations of the ``Hubble Tarantula Treasury Project (HTTP: Unraveling Tarantula's Web)'', an ongoing panchromatic HST survey designed to resolve and characterize the stellar populations in the Tarantula Nebula down to the sub-solar mass regime ($<0.5\, {\rm M_\odot}$). HTTP is collecting deep observations of the Tarantula Nebula in the NUV (F257W and F336W), optical (F555W and F658N), and NIR (F110W and F160W). These data are combined with deep archival observations of the region in the F775W filter.  

The analysis of the data acquired this far indicates that the reddening is highly variable across the region, with the young UMS stars being on average less extinguished by dust than the old RGB stars. This could be a sign that 30~Dor is on the nearer side of the LMC disk. Because the extension of 30~Dor is likely negligible with respect to the LMC depth, the majority of the evolved stars could be beyond 30 Dor, and therefore their luminosity would be extinguished by the dust associated with the Tarantula Nebula. 

We find that stellar populations of different ages have very different spatial distributions. While the intermediate- and old-age stars have had time to diffuse in the disk of the LMC, the younger stars are still associated with their birth sites. In particular the majority of the UMS stars are associated with R136, the NE clump and Hodge~301. 

An inspection of the spatial distribution of the PMS stars reveals the presence of several small clustered systems. The majority of these systems are located between R136 and the northern corner of the area covered by our survey, along the ridge of an X-ray emitting bubble. We speculate that these systems may be minor episodes of star formation that occurred during the fragmentation of the giant molecular cloud that formed R136. Our preliminary analysis confirms that the region has actively formed stars for $\sim 20$ and possibly more Myr. 
Once the NUV and optical data are available, we will use a Bayesian SED-fitting algorithm to derive the star-by-star reddening correction and constrain the age of the various sites of recent star formation. A comparison of our data with synthetic CMDs will allow us to better assess how star formation occurred and is propagating in the Tarantula Nebula, and at which star formation rate.

The uniform coverage and the broad selection of filters of HTTP will benefit the community, for example: 
\begin{enumerate}
\item    H$\alpha$ and near IR images will make it possible to identify emission-line stars such as Be stars, RGB and SGB stars with stellar winds, Herbig Ae/Be stars and low mass PMS stars. \item Super star clusters such as R136 are often considered present-day counterparts of forming globular clusters. HTTP offers an observational snapshot of the early evolutionary phases of these systems and new information on how they relate to their surroundings during early evolutionary phases. \item HTTP will yield the richest and most homogenous sample of moderately metal-poor PMS stars of different ages and masses. SEDs and mass-accretion rates, which can be derived from this dataset, will provide new and much needed constraints on the evolutionary models of PMS stars. \item HTTP can be used to measure the scale at which UV radiation from hot massive stars affects the evolution of low-mass accreting PMS stars.  
\end{enumerate}

HTTP is a treasury program and all the data are immediately available to the public and can be downloaded from the MAST archive.  In addition we will deliver a series of high-level products to the community, including: 
 \begin{enumerate}
 \item A unified star catalog for all the filters (F275W, F336W, F555W, F658N, F775W, F110W, and F160W). For each filter we will provide an average flux and an error in flux, based on multiple independent observations. 
\item Artificial-star tests done with stars inserted along the fiducial sequences. This will allow users to locally quantify the level of completeness of the catalog and the extent to which crowding may have broadened the sequences. 
\item Co-registered stacked images of the field in all filters, with calibrated WCS headers. 
\item Maps of differential reddening and the star formation history of the field. 
\item A catalog with the properties of all the clusters and associations.
 \end{enumerate}

At the moment, only few large ground-based telescopes ($6{\rm m}+$) can surpass the sharpness of Hubble's data at certain wavelengths and on a very limited field of view. In the near future ALMA and JWST will have  spatial resolution and sensitivity similar to HST, but because of the different wavelength coverage, they will probe complementary, but distinct stellar populations and evolutionary stages. In particular JWST and ALMA will be able to identify the most dust-embedded young stellar objects, highlighting where the most massive stars are now forming. In summary, once combined with the our ongoing investigations (i.e. stellar dynamics, PI Lennon; gas dynamics, PI Gallagher) and studies at other wavelengths, HTTP will provide a complete high-resolution picture of the complex interplay between stars, gas, dust and stellar feedback in a bursting regime. HTTP will become the definitive catalog of the field and it will provide a unique way to cross-identify objects in multiple studies, serving as a touchstone for all future works on 30~Dor in particular, and on starbursts in general.
 
\acknowledgments
The authors are grateful to Zolt Levay for his work on the images shown in Figures~1 and 3. MT and MC have been partially founded by contracts ASI I009/10/0, PRIN-INAF-2010 and PRIN-MIUR-2010-11. Support for programs GO-12499 and GO-12939 was provided by NASA through grants from the Space Telescope Science Institute, which is operated by the Association of Universities for Research in Astronomy, Inc., under NASA contract NAS 5-26555. EKG acknowledges support by the Collaborative Research Center ``The Milky Way System'' (SFB 881) of the German Research Foundation (DFG), particularly by subproject B5. 

 \bibliographystyle{apj}

\begin{thebibliography}{}

\bibitem[Andersen et al.(2009)]{andersen09}
Andersen, M., Zinnecker, H., Moneti, A., McCaughrean, M.J., Brandl, B., Brandner, W., Meylan, G., Hunter, D. 2009, ApJ, 707, 1347 

\bibitem[Anderson \& Bedin(2010)]{anderson10}
Anderson, J., Bedin, L.R. 2010, PASP, 122, 1035

\bibitem[Anderson et al.(2008)]{andreson08}
Anderson, J., Sarajedini, A., Bedin, L.R., King, I.R., Piotto, G., Reid, I.N., Siegel, M., Majewski, S.R., Paust, N.E.Q., Aparicio, A., Milone, A.P., Chaboyer, B., Rosenberg, A. 2008, AJ, 135, 2055 

\bibitem[Anderson \& King(2006)]{anderson06}
Anderson, A., \& King, J.R. 2006, STSCI Inst. Sci. Rep ACS 2006-01 (Baltimore: STScI)

\bibitem[Annibali et al.(2003)]{annibali03}
Annibali, F., Greggio, L., Tosi, M., Aloisi, A., Leitherer, C. 2003, AJ, 126, 2752 

\bibitem[Annibali et al.(2009)]{annibali09}
Annibali, F.,  Tosi, M., Monelli, M., Sirianni, M., Montegriffo, P., Aloisi, A., Greggio, L. 2009, AJ, 138, 169

\bibitem[Aparicio et al.(1996)]{aparicio96}
Aparicio A, Gallart C, Chiosi C, Bertelli G. 1996, ApJ, 469,L97

\bibitem[Baggett \& Anderson(2012)]{baggett12}
Baggett, S., \& Anderson, J. 2012, ``WFC3/UVIS Sky Backgrounds'' WFC3-ISR 2012-12

\bibitem[Bate (2009)]{bate09}
Bate M. R., 2009, MNRAS, 392, 590

\bibitem[Bonnell et al.(2011)]{bonnell11}
Bonnell, I.A., Smith, R.J., Clark, P.C., Bate, M.R. 2011, MNRAS, 410, 2339

\bibitem[Brandner et al.(2001)]{brandner01}
Brandner, W., Grebel, E.K., Barb\'a, R.H., Walborn, N.R., Moneti, A. 2001, AJ, 122, 858 

\bibitem[Brandl et al.(2004)]{brandl04}
Brandl, B.R., et al. 2004, ApJS, 125, 188

\bibitem[Chen et al.(2006)]{chen06}
Chen, Y., Wang, Q.D., Gotthelf, E.V., Jiang, Bing, Chu, Y.-H., Gruendl, R. 2006, ApJ, 651, 237

\bibitem[Chu et al.(1994)]{chu}
Chu, Y. -H., Kennicutt, R.C., Jr 1994, ApJ, 108, 1696

\bibitem[Cignoni et al.(2010)]{cignoni2010}
Cignoni, M., Tosi, M., Sabbi, E., Nota, A., Degl'Innocenti, S., Prada Moroni, P. G., Gallagher, J. S. 2010, ApJ, 712, L63

\bibitem[Cioni et al.(2011)]{cioni11}
Cioni, M.-R. et al. 2011, A\&A, 527, 116

\bibitem[Crowther et al. (2010)]{crowther10}
Crowther, P.A.,  et al. 2010, MNRAS, 408, 731

\bibitem[Dekel \& Silk(1986)]{dekel86}
Dekel, A., Silk, J. 1986, ApJ, 303, 39

\bibitem[de la Caille(1755)]{caille55}
de la Caille, N.L. 1755, Mem. Acad. annee 1755, 194

\bibitem[De Marchi et al.(2011)]{demarchi11}
De Marchi, et al. 2011, ApJ, 739, 27

\bibitem[De Marchi et al.(2010)]{demarchi2010}
De Marchi, G.,  Panagia, N., Romaniello, M. 2010, ApJ, 715, 1

\bibitem[Douglas et al.(2010)]{douglas10}
Douglas, L.S., Bremer, M.N., Lehnert, M.D., Stanway, E.R. 2010, MNRAS, 409, 1155

\bibitem[Dressel(2011)]{dressel11}
Dressel, L. 2011, ``Wide Field Camera 3 Instrument Handbook, Version 4.0'' (Baltimore: STScI)

\bibitem[Elmegreen \& Lada(1977)]{elmegreen77}
Elmegreen, B.G., Lada, C.J. 1977, ApJ, 214, 725

\bibitem[Evans et al.(2010)]{evans10}
Evans, C.J., et al. 2010, ApJ, 715, L74

\bibitem[Evans et al.(2011)]{evans11}
Evans, C.J. et al. 2011, A\&A, 530, A108

\bibitem[Feast(1953)]{feast53}
Feast, M.W. 1953, The Observatory, 73,  255 

\bibitem[Ferguson \& Babul(1998)]{ferguson98}
Ferguson, H., Babul, A. 1998, MNRAS, 296, 585

\bibitem[Feitzinger et al.(1980)]{feitzinger80}
Feitzinger, J. V., Schlosser, W., Schmidt-Kaler, T., Winkler, C. 1980, A\&A, 84, 50

\bibitem[Gallart, Aparicio, \& Vilchez(1996)]{gallart96}
Gallart, C., Aparicio, A., \& Vilchez, J. M. 1996, AJ, 112, 1928

\bibitem[Gilmozzi et al.(1994)]{gilmozzi94}
Gilmozzi, R., Kinney, E.K., Ewald, S.P., Panagia, N., Romaniello, M. 1994, ApJ, 435, L43

\bibitem[Gouliermis et al.(2012)]{gouliermis2012}
Gouliermis, D. A., Schmeja, S., Dolphin, A. E.; Gennaro, M., Tognelli, E., Prada Moroni, P. G. 2012, ApJ, 748, 64

\bibitem[Grebel(1997)]{grebel97}
Grebel, E. K 1997, A\&A, 317, 448

\bibitem[Grebel \& Chu(2000)]{grebel00}
Grebel, E.K., Chu, Y.-H., 2000, AJ, 111, 787

\bibitem[Haschke et al.(2011)]{haschke11}
Haschke, R., Grebel, E.K., Duffau, S. 2011, AJ, 141, 158

\bibitem[Harris et al. (2004)]{harris04}
Harris, J., Calzetti, D., Gallagher, J. S., Smith, D. A., \& Conselice, C. J. 2004, ApJ, 603, 503

\bibitem[Harris \& Zaritsky(2001)]{harris01}
Harris, J, \& Zaritsky, D. 2001, ApJS, 136, 25

\bibitem[Heckman et al.(2004)]{heckman04}
Heckman, T.M., Kauffmann, G., Brinchmann, J., Charlot, S., Tremonti, C.,White, S.D.M. 2004, ApJ, 613, 109

\bibitem[Heckman(2005)]{heckman05}
Heckman, T.M. 2005, ``tarbursts: From 30 Doradus to Lyman Break Galaxies'', Held in Cambridge, UK, 6-10 September 2004. Edited by R. de Grijs and R.M. Gonz‡lez Delgado, ASSL, 329, 3

\bibitem[Helling et al.(2000)]{helling00}
Helling, C., Winters, J.M., Sedlmayr, E. 2000, A\&A, 358, 651

\bibitem[Herbig(1962)]{herbig62}
Herbig, G. H. 1962, ApJ, 135, 736

\bibitem[Herschel(1847)]{herschel47}
Herschel, J.F.W. 1847, ``Results of astronomical observations made during the years 1834, 5, 6, 7, 8, at the Cape of Good Hope; being the completion of a telescopic survey of the whole surface of the visible heavens, commenced in 1825'', London, Smith, Elder \& C.

\bibitem[Holtzman et al.(1999)]{holtzman99}
Holtzman, J.A., et al. 1999, AJ, 118, 2262

\bibitem[Hunt \& Hirashita(2009)]{hunt09}
Hunt, L.K., Hirashita, H. 2009, A\&A, 507, 1327

\bibitem[Hunter \& Elmegreen(2006)]{hunter06}
Hunter, D.A., Elmegreen, B.G. 2006, ApJS, 162, 49

\bibitem[Kennicutt(1984)]{kennicutt84}
Kennicutt, R.C.Jr. 1984, ApJ, 287, 116

\bibitem[Kennicutt \& Hodge(1986)]{kennicutt86}
Kennicutt, R.C., \& Hodge, P.W. 1986, ApJ, 306, 130

\bibitem[Kennicutt \& Evans(2012)]{kennicutt12}
Kennicutt, R.C., Evans, N.J. 2012, ARA\&A, 50, 531

\bibitem[Konstantopoulos et al.(2009)]{konstantopoulos09}
Konstantopoulos, I.S., Bastian, N., Smith, L.J., Westmoquette, M.S., Trancho, G., Gallagher, J.S. III 2009, ApJ, 701, 1015

\bibitem[Lada \&Lada(2003)]{lada03}
Lada, C.J., \& Lada, E. 2003, ARA\&A, 41, 57

\bibitem[Lee et al.(2007)]{lee07}
Lee, J.C., Kennicutt, R.C., Funes, S.J., Jos\'e G., Sakai, S., Akiyama, S. 2007, ApJ, 671, L113

\bibitem[Lee et al.(2009)]{lee09}
Lee, J.C., et al. 2009, ApJ, 706, 599

\bibitem[Leitherer(1998)]{leitherer98}
Leitherer, C. 1998, ``Stellar astrophysics for the local group : VIII Canary Islands Winter School of Astrophysics''. Edited by A. Aparicio, A. Herrero, and F. Sanchez. Cambridge; New York: Cambridge University Press, 1998., p.527

\bibitem[Mac Low \& Ferrara(1999)]{maclow99}
Mac Low, M.-M., Ferrara, A. 1999, ApJ, 513, 142

\bibitem[MacKenty \& Smith(2012)]{mackenty12}
MacKenty, J.W., Smith, L.J. 2012, ``CTE White Paper'', Space Telescope Science Institute, http://www.stsci.edu/hst/wfc3/ins\_performance/CTE/CTE\_White\_Paper.pdf

\bibitem[Madau et al.(1996)]{madau96}
Madau, P., Ferguson, H.C., Dickinson, M. E., Giavalisco, M., Steidel, C. C., Fruchter, A., 1996, MNRAS, 283, 1388

\bibitem[Mas-Hesse, \& Kunth(1999)]{mashesse99}
Mas-Hesse, J. M., Kunth, D. 1999, A\&A, 349, 765

\bibitem[Massey, et al.(1998)]{massey98}
Massey, P., Lang, C.C., Degioia-Eastwood, K.,  Garmany, C.D. 1995, ApJ, 438, 188

\bibitem[McQuinn et al.(2012)]{mcquinn12}
McQuinn, K.B.W., Skillman, E.D., Dalcanton, J.J., Cannon, J.M., Dolphin, A.E., Holtzman, J., Weisz, D.R., Stark, D., Weisz, D., Williams, B.F. 2010, ApJ, 724, 49

\bibitem[McQuinn et al.(2010)]{mcquinn10}
McQuinn, K.B.W., Skillman, E.D., Cannon, J.M., Dalcanton, J.J., Dolphin, A.E., Holtzman, J., Weisz, D.R., Williams, B.F. 2012, ApJ, 759, 77

\bibitem[Meaburn (1984)]{meaburn84}
Meaburn, J. 1984, MNRAS, 211, 521

\bibitem[Meixner et al(2006)]{meixner06}
Meixner, M. et al. 2006, AJ, 132, 2268

\bibitem[Meixner et al(2010)]{meixner10}
Meixner, M. et al. 2010, A\&A, 518, L71

\bibitem[Meurer et al.(1997)]{meurer97}
Meurer, G.R., Heckman, T.M., Lehnert, M.D., Leitherer, C., Lowenthal, J. 1997, AJ, 114, 54

\bibitem[Moffat(1982)]{moffat82}
Moffatt, A. F.J. 1982, JRASC, 76, 323

\bibitem[Nikolaev et al.(2004)]{nikolaev04}
Nikolaev S., Drake, A.J., Keller, S.C., Cook, K.H., Dalal, N., Griest, K., Welch, D.L., Kanbur, S.M. 2004, ApJ,  601, 260 

\bibitem[O'Connell \& Mangano(1978)]{oconnel78}
O'Connell, R. W., Mangano, J. J. 1978, ApJ, 221, 62

\bibitem[O'Connell et al.(1995)]{oconnel95}
O'Connell, R. W., Gallagher, J. S., III, Hunter, D. A., Colley, W. N. 1995, ApJ, 446, L1

\bibitem[O'Connell(2005)]{oconnell05}
O'Connell, R.W. 2005, SSL 329, Starbursts: From 30 Doradus to Lyman Break Galaxies, ed. R. de Grijs \& R.M. Gonz\`alez Delgado (The Netherlands: Springer), 333

\bibitem[Oey et al.(2003)]{oey03}
Oey, M.S., Parker, J.S., Mikles, V.J., Zhang, X. 2003, AJ, 126, 2317

\bibitem[Panagia et al.(2000)]{panagia00}
Panagia, N., Romaniello, M., Scuderi, S., Kirshner, R.P. 2000, ApJ, 539, 197

\bibitem[Panagia et al.(1991)]{panagia91}
Panagia, N., Gilmozzi, R., Macchetto,  F., Adorf, H.M., Kirshner, R.P. 1991, ApJ, 380, L23  

\bibitem[Pei et al. (1999)]{pei99}
Pei, Y. C., Fall, S. M. \& Hauser, M. G., 1999, ApJ, 522, 604

\bibitem[Pietrzy\'nski et al.(2013)]{pietr13}
Pietrzy\'nski, G. et al. 2013, Nature, 495, 76

\bibitem[Romaniello et al.(2002)]{romaniello02}
Romaniello, M., Panagia, N., Scuderi, S., Kirshner, R. P. 2002, AJ, 123, 915

\bibitem[Sabbi et al.(2007)]{sabbi07}
Sabbi, E., Sirianni, M., Nota, A., Tosi, M., Gallagher, J., Meixner, M., Oey, M. S., Walterbos, R., Pasquali, A., Smith, L. J., Angeretti, L 2007, AJ, 133, 44

\bibitem[Sabbi et al.(2012)]{sabbi12}
Sabbi, E., Lennon, D.J., Gieles, M., de Mink, S.E., Walborn, N.R., Anderson, J., Bellini, A., Panagia., N., van der Marel, R., Mai\'z Apellan\'iz, J. 2012, ApJ, 754, L37

\bibitem[Sabbi et al.(2008)]{sabbi08}
Sabbi, E., Sirianni, M., Nota, A., Tosi, M., Gallagher, J., Smith, L. J., Angeretti, L., Meixner, M., Oey, M. S., Walterbos, R., Pasquali, A. 2008, AJ, 135, 173

\bibitem[Searle et al.(1973)]{searle73}
Searle, L., Sargent, W.L.W., Bagnuolo, W.G. 1973, ApJ, 179, 427

\bibitem[Schaerer, Contini, \& Kunth(1999)]{schaerer99}
Schaerer, D., Contini, T., Kunth, D. 1999, A\&A, 341, 399

\bibitem[Skrutskie  et al.(2006)]{skrutskie06}
Skrutskie, M.F. et al. 2006, AJ, 131, 1163

\bibitem[Shapley et al.(2003)]{shapley03}
Shapley, A.E., Steidel, C.C., Pettini, M., Adelberg, K.L. 2003, ApJ, 588, 65

\bibitem[Stinson et al.(2007)]{stinson07}
Stinson, G.S., Dalcanton, J.J., Quinn, T., Kaufmann, T., Wadsley, J. 2007, ApJ, 667, 170 

\bibitem[Thornley et al.(2000)]{thornley00}
Thornley, M. D., Schreiber, N. M. F., Lutz, D., Genzel, R., Spoon, H. W. W., Kunze, D. 2000, ApJ, 539, 641

\bibitem[Tolstoy, Hill, \& Tosi(2009)]{tolstoy09}
Tolstoy, E., Hill, V., \& Tosi, M. 2009, ARA\&A, 47, 371

\bibitem[Tolstoy(1996)]{tolstoy96}
Tolstoy E. 1996, ApJ, 462, 684

\bibitem[Tosi et al.(1991)]{tosi91}
Tosi M, Greggio L, Marconi G, Focardi P. 1991,AJ,102,951

\bibitem[Tosi et al.(1989)]{tosi89}
Tosi, M., Greggio, L., Focardi, P. 1989, Ap\&SS, 156, 295

\bibitem[Townsley et al.(2006)]{townsley06}
Townsley, L.K., Broos, P.S., Feigelson, E.D., Brandl, B.R., Chu, Y.-H.; Garmire, G.P., Pavlov, G.G. 2006, AJ, 131, 2140
	
\bibitem[Tremonti et al.(2001)]{tremonti01}
Tremonti, C. A., Calzetti, D., Leitherer, C., Heckman, T. M. 2001, ApJ, 555, 322	

\bibitem[Vanhala \&  Cameron(1998)]{vanhala98}
Vanhala, Harri A. T.; Cameron, A. G. W. 1998, ApJ, 508, 291	

\bibitem[Walborn (1973)]{walborn73}
Walborn, N. R. 1973, ApJ, 182, L21	

\bibitem[Walborn \& Blades(1997)]{walborn97}
Walborn, N., Blades, J.C. 1997, ApJS, 112, 457

\bibitem[Walborn et al.(1999)]{walborn99}
Walborn, N.R., Barb\'a, R.H., Brandner, W., Rubio, M., Grebel, E.K., Probst R.G. 1999, AJ, 117, 225

\bibitem[Walborn et al.(2002)]{walborn01}
Walborn, N.R., Ma\'iz Apell\'aniz, J., Barba, R.H. 2002, AJ, 124, 1601

\bibitem[Walborn et al. (2013)]{walborn13}
Walborn, N.R., Barb\`a, R., \& Sewillo, M 2013, AJ, (accepted), astro-ph/1302.3533 

\bibitem[Wang \& Helfand(1991)]{wang91}
Wang, Q, \& Helfand, D. J. 1991, ApJ, 370, 541

\bibitem[Weigelt \& Baier(1985)]{weigelt85}
Weigelt, G., \& Baier, G. 1985, A\&A, 150, L18

\bibitem[Withmore, Chandar, \& Fall(2007)]{withmore07}
Whitmore, B.C., Chandar, R., Fall, S.M. 2007, AJ, 133, 1067 

\bibitem[Zaritsky(1999)]{zaritsky99}
Zaritsky, D. 1999, AJ, 118, 2824

\end{thebibliography}

\end{document}